# High Power, Continuous-wave Supercontinuum Generation in Highly Nonlinear Fibers Pumped with High Order, Cascaded Raman Fiber Amplifiers

VISHAL CHOUDHURY[1,2], S. ARUN[1,2], ROOPA PRAKASH[1], V. R. SUPRADEEPA[1,*]

[1]Center for Nano Science and Engineering, Indian Institute of Science, Bangalore 560012, India
[2]These authors contributed equally to this paper
*Corresponding author: supradeepa@iisc.ac.in

A novel method for efficient generation of high power, equalized continuous-wave supercontinuum source in an all conventional silica fiber architecture is demonstrated. Highly nonlinear fiber (HNLF) is pumped in its anomalous dispersion region using a novel, high power, L-band laser. The L-band laser encompasses a 6th order cascaded Raman amplifier which is pumped with a high power Ytterbium doped fiber laser and amplifies a low-power, tunable L-band seed source. The supercontinuum generated 35W of power with ~40% efficiency. The Supercontinuum spectrum was measured to have a high degree of flatness of better than 5 dB over 400 nm of bandwidth (1.3–1.7µm, limited by spectrum analyzer range) and a power spectral density in this region of >50 mW/nm. The extent of the SC spectrum is estimated to be upto 2 µm

## INTRODUCTION

High power continuous-wave (CW) Supercontinuum (SC) fiber lasers are widely investigated light sources for applications such as spectroscopy, gas sensing, LIDAR, Test and Measurement and Optical communications [1-4]. Through filtering of supercontinuum sources, lasers are enabled in wavelength regions inaccessible to conventional rare-earth doped fiber lasers. Generation of CW Supercontinuum requires pumping high power in the anomalous dispersion region of non-linear media such as highly nonlinear fibers or photonic crystal fibers. Through a combination of multiple nonlinear effects such as modulational instability, soliton effects, four-wave mixing and stimulated Raman scattering, a broad optical spectra can be generated [5]. Conventionally, photonic crystal fibers (PCF) are used as the nonlinear medium [6, 7]. This is because Yb lasers with emission near 1 µm region have shown the highest power scalability over the last two decades [8]. In contrast to conventional silica fibers whose zero dispersion wavelength is at longer wavelengths, the zero dispersion wavelength (ZDW) of PCFs can be designed to be near 1µm. CW SC with power as high as 29W has been reported in PCFs with Yb fiber laser pumping [9]. Due to inherent material properties of silica, conventional silica fibers have ZDW beyond 1.3µm. In addition, the quality of the supercontinuum generated is enhanced with increasing nonlinear coefficient of the fiber. If the nonlinear coefficient of the fiber has to be increased, the mode area of the silica fibers needs to be reduced (as in the case of highly nonlinear fiber (HNLF)). This plays an additional role in pushing the zero dispersion wavelength to even longer wavelengths such as beyond 1.5µm. However, in contrast to the 1µm wavelength region, scalable, high power, high brightness (single mode) sources are lacking in this wavelength region. This is a significant limitation since using conventional silica fibers has several advantages over PCFs. Apart from being difficult in fabricating as compared to the conventional fibers, difficulties in fusion splicing PCFs with other conventional silica based fibers is often encountered. As a result, power into the PCF is often coupled using free-space optics preventing an all-fiber, fusion spliced architecture. Additional coupling losses in this configuration results in less power coupled from the pump laser to the nonlinear medium which can potentially reduce performance and efficiency. Even based on cost considerations, PCFs used for nonlinear applications tend to be more expensive than nonlinear conventional silica fibers (HNLF). Due to these reasons, HNLF has been utilized by several groups for CW SC generation [10-12]. In these results, the supercontinuum was pumped with a Yb doped fiber laser around 1.05-1.1µm, in the normal dispersion region and relied on the use of cascaded Raman scattering. Here, the power is transferred from the input wavelength to longer wavelengths in the anomalous dispersion region through a series of Raman shifts following which the supercontinuum generation occurs. In silica fibers, the peak of the Raman gain exists around a frequency shift of ~13 THz. This necessitates several cascaded Raman frequency shifts to bridge the pump wavelength in the Yb emission region to the zero dispersion wavelength region (beyond 1.3 µm). It has been observed that pumping in normal dispersion additionally enhances the flatness of the spectrum [10, 13]. The mechanism for this and technique to utilize this here is discussed in later sections.

A key requirement to generate supercontinuum by pumping conventional silica fibers in the normal dispersion region is to enable cascaded Raman conversion [14, 15]. In [10], high power from the output of a Yb doped fiber laser is directly pumped into the HNLF and no additional steps are taken to specifically enable efficient cascaded Raman scattering. Owing to this reason, the efficiency of the supercontinuum generation is relatively low (19W output power with 14% efficiency w.r.t. Yb fiber laser input). In addition, it has been observed that, in the absence of steps taken to enable preferential forward Raman scattering, there is a potential for destabilization and damage to the input Yb fiber laser due to backward propagating Raman shifted light. In [11, 12], a cascaded Raman resonator was utilized. This involves a resonator constituting a series of nested cavities at all the intermediate Raman Stokes wavelengths. These cavities are made through the use of a series of high reflectivity Fiber Bragg gratings at all the intermediate Stokes wavelengths in the input and output side. A low reflectivity output coupler at the required output wavelength terminates the cascaded Raman conversion. The fiber utilized is a low effective area, high nonlinearity fiber specifically designed for high Raman gain, referred to as Raman fiber. At the output of the cascaded Raman resonator, most of the light is in the required output wavelength. However, the output power was relatively low (< 5W) and the supercontinuum demonstrated a substantial power difference across the wavelengths with the normal dispersion side being ~20-dB lower than the anomalous dispersion side.

Here, we demonstrate a supercontinuum source based on a novel, all-passive cascaded Raman fiber laser architecture we have proposed recently which overcomes both these problems. By seeding the cascaded Raman conversion at intermediate wavelengths, high conversion efficiencies are enabled. This efficiency benefit is also transferred to the output of the supercontinuum. Further, in contrast to the cascaded Raman resonator based system, the proposed system allows for the power to be distributed across the intermediate Stokes wavelengths in the normal dispersion region instead of confining the power within a single output wavelength at the operating power level. This enhances the spectral quality of the supercontinuum with similar spectral flatness occurring on either side of the zero-dispersion wavelength. Here, we demonstrate a 35W supercontinuum source extending from 1.3 to 2.0μm with ~40% efficiency and a flatness level of better than 5-dB over a bandwidth of at least 400nm (measurement limited by wavelength range of the optical spectrum analyzer).

## EXPERIMENT & RESULTS

Figure 1 shows the experimental setup. In our experiment, we use a recently proposed 6th order cascaded Raman wavelength convertor [16]. The system is grouped into 3-primary constituents (a), (b) and (c) based on function. The first constituent (a) is a 1117nm, Yb doped fiber laser based on cladding pumped architecture using 976nm pump diodes. The laser uses 15m of 5/130 Yb doped gain fiber (Nufern) and 3 tentatively '50'W laser diode modules. A maximum output power of ~90W is obtained from this laser. In addition, a home-built tunable, L-band seed source is coupled into the unused signal port of the pump combiner. The ring laser is based on Erbium-Ytterbium co-doped fiber with an intracavity thin-film filter for tuning and can tune from 1565nm to 1590nm. This enables a convenient method to couple signal power without the need for additional components. In (b), a 6th order cascaded Raman convertor is implemented. A 1117/1480 nm fused fiber WDM couples a small fraction of the 1117nm light due to non-ideal isolation into a separate path where a conventional cascaded Raman resonator is implemented. Here, 120m of Raman fiber with ~12sq micron effective area is utilized. The Raman input and output grating sets (RIG) and (ROG) are a series of high reflecting (>99%) fiber Bragg gratings at all the intermediate Stokes wavelengths. A low reflectivity output coupler (7%) in the ROG at 1480nm terminates the cascade. At the output of the cascaded Raman resonator, in addition to output power at 1480nm, there is a small fraction of power at all the intermediate Raman Stokes wavelengths. These are all coupled back into the primary path through the WDM as shown in the figure, followed by 250m more of Raman fiber. The presence of seeding of the intermediate Stokes components and avoiding the cascaded Raman resonator assembly in the primary path enable efficient, cascaded Raman conversion. The seed wavelengths generated in the cascaded Raman resonator together with the input L-band laser, a 6th order cascaded Raman amplifier is implemented. Power is transferred to the L-band signal from the Yb doped fiber laser pump. Additional details on this architecture can be found in [15, 16]. In (c), we have the supercontinuum generation stage. 100m of off the shelf HNLF acquired from OFS Denmark was utilized. The specifications of the HNLF from the vendor are a non-linear coefficient ($\gamma$) of 11.3 W$^{-1}$km$^{-1}$ and ZDW of 1553 nm. The HNLF was wound on an aluminum spool of ~9cm diameter to conductively cool the fiber during operation. The tunability of the L-band seed was utilized to investigate the changes in the supercontinuum as the wavelength is changed.

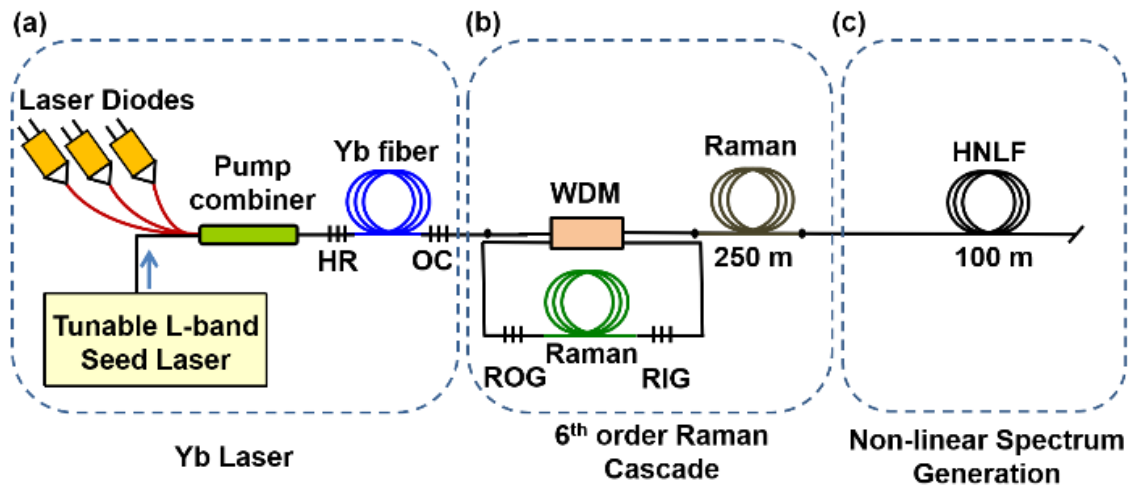

Fig. 1. Experimental setup. The 1117 Yb laser in (a) is used as the primary pumping source for SRS. The Raman Input Grating (RIG) and Raman Output Grating (ROG) (b) sets form the 5th order cavity which seeds the five Raman Stokes while the low power L- band laser from (a) acts as the seed for the 6th Raman Stokes. SC generation (c) takes place as the 6th order Stokes goes into the HNLF.

Figure 2 shows the input output power relations of the supercontinuum source for input seed at 1570nm. An output power of over ~35W was measured limited by the pump power in the Yb-laser. The power conversion efficiency w.r.t. the 1117 nm Yb laser to the SC is ~40%. The high conversion efficiency in the SC (over twice the output power and thrice the efficiency compared to the previous result with HNLF) in our architecture is a consequence of using a fully seeded cascaded Raman fiber laser.

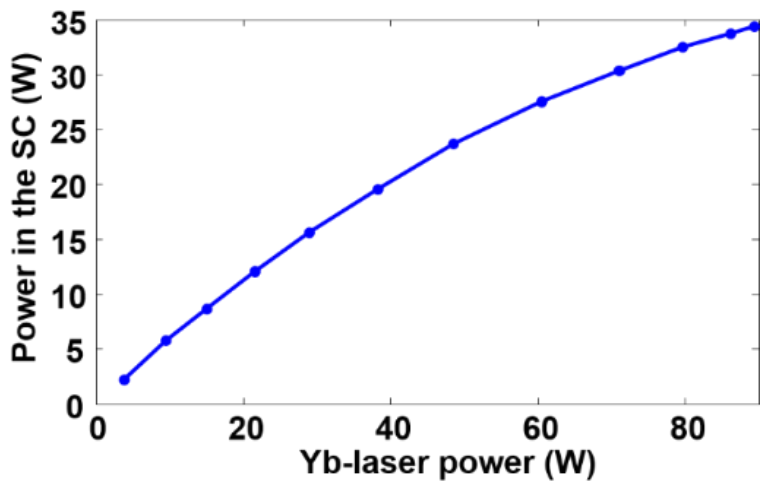

Fig. 2. Output power of the SC w.r.t. the input 1117 nm Yb laser power.

Figure 3 shows the output spectrum of the supercontinuum at the full power level of 35W for the seed laser at 1570nm. The output spectrum was measured by an optical spectrum analyzer with a limited wavelength range until 1700nm. As shown in the figure, a broadband and very flat supercontinuum is obtained. Flatness of better than 5-dB was obtained over a band of ~400nm from 1300nm to 1700nm. This is limited by the measurement range of the OSA and we anticipate the flat region to further extend beyond 1700nm. The SC spectra was measured to be stable over 10minutes of continuous operation and the temperature of the HNLF spool was observed to not increase beyond 10 degrees higher than the ambient temperature. If we notice the short-wavelength side of the supercontinuum, a sharp cutoff exists around 1300nm. Since the short-wavelength cutoff arises due to a four-wave mixing process between the pump and the long-wavelength cutoff, from the value of the pump wavelength (1570nm) and the short-wavelength cutoff (~1300nm), we anticipate that the long-wavelength cutoff of the supercontinuum is at ~1980nm. The true extent of the supercontinuum source is thus anticipated to be ~700nm. Additional measurements with a long-wavelength spectrum analyzer is necessary to verify this. In the region of accurate measurement, we could obtain over 50mW/nm of spectral density from 1300 to 1700nm. This is the highest spectral density to the best of our knowledge in a broadband CW supercontinuum source based on an all fiber architecture. The high spectral density can enable on filtering, a wide variety of tunable CW lasers at different wavelength bands with substantial power. Recently, we had proposed a different CW supercontinuum source based on SMF and using Random Raman Fiber lasers [17]. The bandwidth obtained there was higher (~1000nm) and the power level and efficiency was comparable, however, the spectral flatness was poor with over 20-dB variations in amplitude across the spectrum. Here, with the use of HNLF and a fully seeded architecture, there has been an enormous improvement in the spectral flatness of the supercontinuum, a very desirable property for these sources.

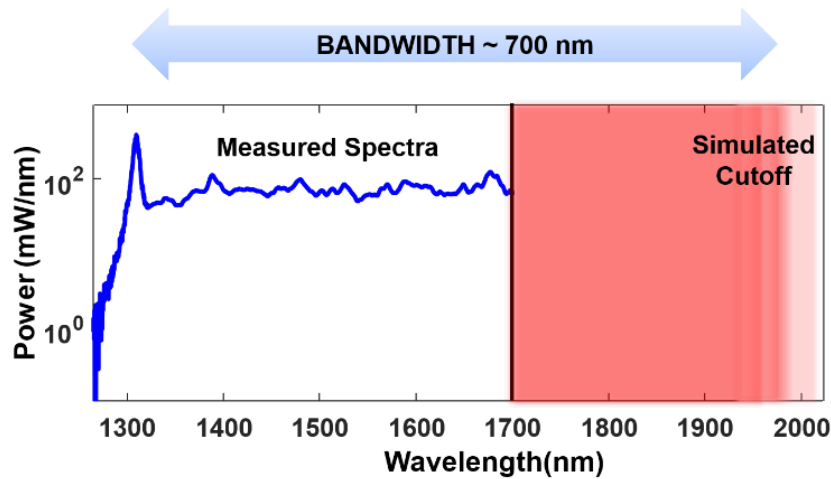

Fig. 3. The measured SC spectra. The experimentally measured part of the spectra starts from ~1300 nm and ends at 1700 nm limited by the wavelength range of the OSA. The actual cutoff of the spectra is determined based on the short wavelength cutoff.

The seed wavelength was continuously tuned across 1565nm to 1590nm to analyze the supercontinuum spectrum. It was observed that the output power level and spectral shape remained relatively unchanged across this range. This is a very desirable attribute since the zero dispersion wavelengths of HNLFs of the same design can vary due to process variations. The robustness in the supercontinuum generation, with regard to the location of the pump wavelength in relation to the zero dispersion wavelength can substantially reduce the specification requirements for the HNLF used in this application.

Looking at the mechanism of the generation process in more detail, the extension of the SC in the anomalous dispersion is driven by the formation of solitons. The time domain fluctuations present in the envelope of the CW laser seeds modulational instability (MI) which results in solitons of different widths and peaks. These pulses undergo spectral broadening towards the longer wavelength side through the phenomenon of Raman Induced Frequency Shift (RIFS). Simulations carried by several groups showed that the solitons when perturbed with third order dispersion give rise to blue shifted dispersive waves called Cherenkov radiation [5, 13, 18]. This phenomenon results in the spectral broadening by the generation of new frequencies in the normal dispersion side of the spectra. These dispersive waves are phase matched with the solitons and thus propagate together along the fiber. Interaction of solitons with other solitons and with the dispersive waves through FWM results in further spectral broadening of the SC.

Due to the contribution of RIFS towards only the longer wavelength region, the portion of the SC in anomalous dispersion always attains much more power spectral density (PSD) than the portion in normal dispersion. This results in an asymmetry in the spectra [11, 12]. The flatness of the SC spectra can be significantly enhanced in presence of a CW pump in the normal dispersion region. Experiments and simulations have shown that FWM conditions between the solitons, the CW pump and the dispersive waves can be attained in optical fibers. As a result, pumping CW lasers in the normal dispersion region seeds the soliton-dispersive wave interaction which increases the PSD in the normal dispersion [10, 13]. This increases the degree of flatness of the spectra. The Raman fiber laser architecture utilized in our experiment allows for a distributed pump scheme which helps in attaining a very high degree of flatness in the spectra. In our system, the intermediate Raman Stokes lines in 1390 nm and 1480 nm bands act as individual CW pumps to seed the generation of dispersive waves in the normal dispersion region. The distributed pump scheme offered by our architecture flattens the spectrum which is not seen when pumped with only one CW laser [11, 12, 19].

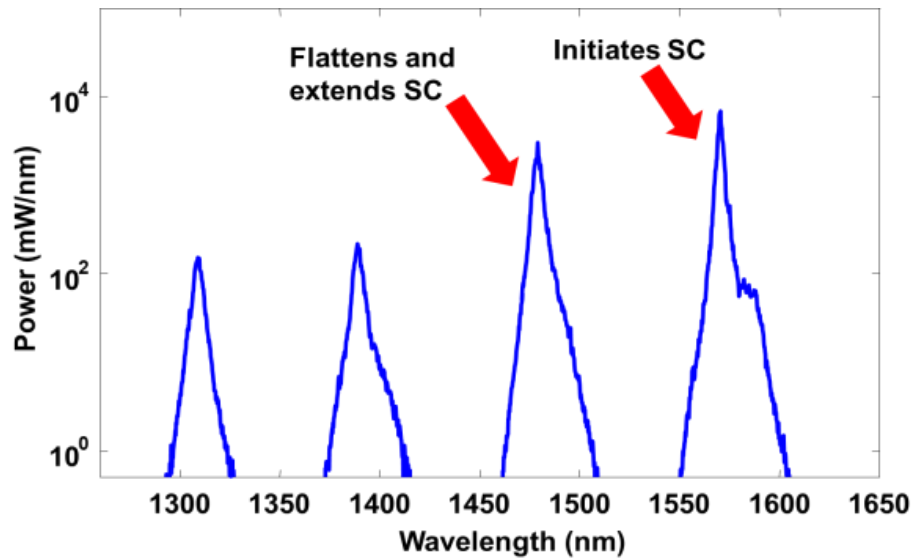

Fig. 4. Spectra at full power before the HNLF. The 1570 band initiates the SC via MI and the remaining bands help seed the soliton-dispersive waves interaction which flattens the SC. Total power input is ~40 W.

The Raman conversion in our system was implemented in such a way (Fig. 4) that an optimum power distribution in the normal dispersion region of the HNLF is ensured. This was obtained through optimizing the length of Raman fiber used for cascaded Raman conversion in the stage prior to the HNLF. As shown in the figure, prior to entering the HNLF, substantial components are seen not only at the 1570nm which initiates the supercontinuum, but also at intermediate Stokes wavelengths of 1480nm, 1390nm and 1310nm. These wavelengths in the normal dispersion region of the HNLF assist in flattening the SC and extending it in the normal dispersion region.

Figure 5 shows the evolution of the supercontinuum with increasing power. In Fig. 5(a), the Raman conversion to 1480 nm band is still taking place and the SC is in a state of initiation. An asymmetry of the supercontinuum between the anomalous and the normal dispersion regions can be seen here. As the power in the 1570 nm band increases, the spectrum also broadens towards the longer wavelength region (Figs 5(b)-5(d)). In Fig 5(b), the Raman conversion into 1480 nm has stopped while 1570 nm is still attaining power and extending the SC. The asymmetry in the spectrum is still present. In Fig 5(c), the SC becomes flatter with 1480 nm and 1570 nm bands attaining the optimum distributed pump state. The asymmetry reduces as power is further enhanced and a broad, symmetric, flat spectrum is observed (Fig 5(d)).

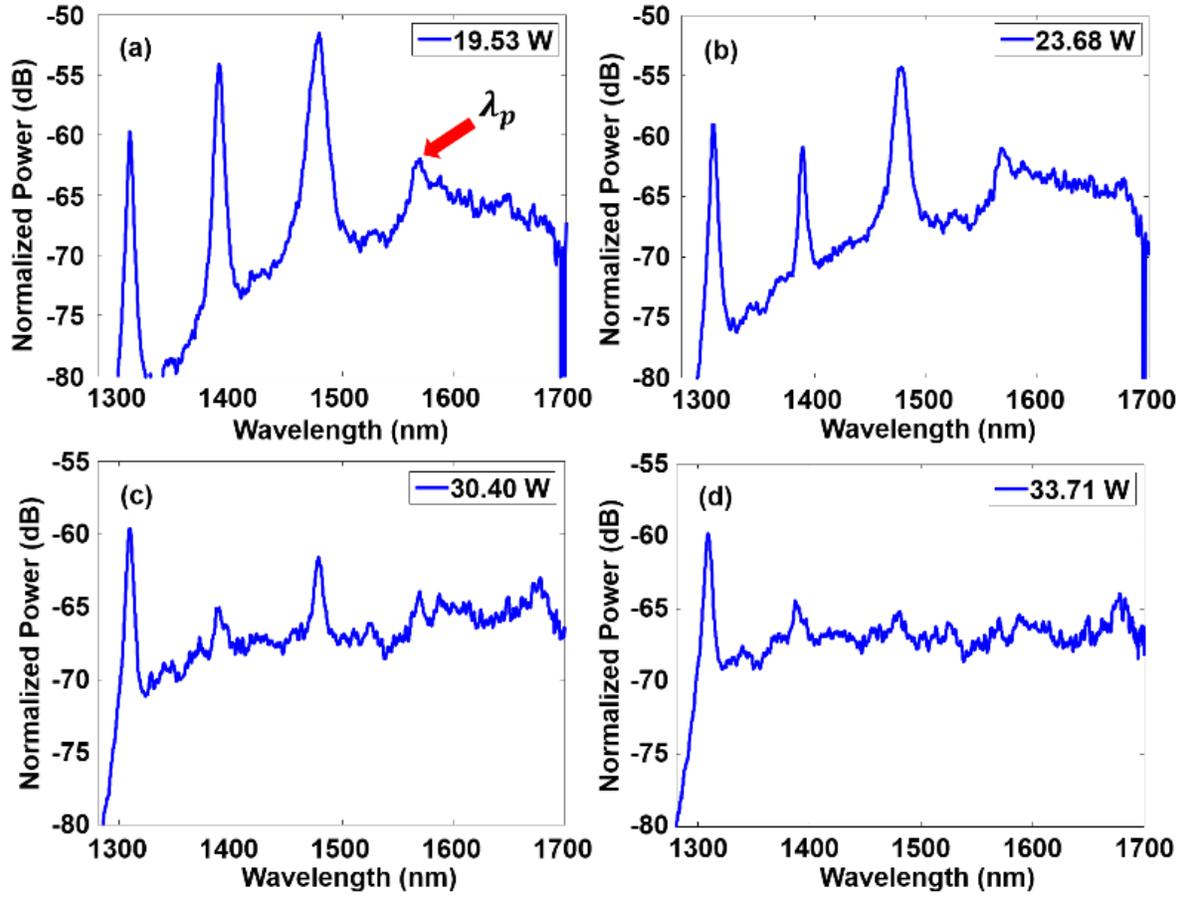

Fig. 5. (a) Spectra when Raman conversion to 1480 nm band is still taking place and the SC is in a state of initiation. In (b) the Raman conversion into 1480nm has stopped while 1570 nm is still attaining power while still extending the SC. In (c), the SC becomes flatter with 1480 nm and 1570 nm bands attaining the optimum distributed pump state which results in a broad flat spectrum with increased power (d).

The span of the SC is essentially limited by the silica attenuation in mid IR. By implementing the degenerate FWM condition $2f_p = f_{short} + f_{long}$, where the two frequencies in the right hand side of the equation indicate the short and long frequency cut-off and the left hand side is the pump frequency, the longer wavelength cutoff ($\lambda_{long}$) was calculated to be around 2μm (~1980nm). This is expected as the silica attenuation drastically increases after 2μm (>20 dB/km). In Fig 5, we see that near the short wavelength cutoff there is also a reasonable fraction of unconverted light from the Raman cascade at 1310nm. This interferes with the accurate evaluation of the short wavelength cutoff. To overcome this, data processing of different spectra as the supercontinuum was evolving was utilized to remove the 1310nm component. We made use of the fact that line-broadening of the 1310 nm component is small in HNLF due to high dispersion at this wavelength and any change in line-shape is by the addition of the supercontinuum spectra underneath it. The short wavelength cutoff was estimated to be ~1300nm and this provides an estimate of long-wavelength cutoff at ~1980nm.

## SUMMARY

We have demonstrated a high efficiency, high power, continuous-wave Supercontinuum source using conventional silica based fibers in an all fiber, fusion-spliced architecture. From a systems perspective, this system does not have free space alignment or bulk laser limitations. Our system provided a pump limited output power of ~35W with an efficiency of ~40% w.r.t. output power of Yb fiber laser pump source. The output spectra was measured to have a high spectral flatness of 5-dB over 400nm of bandwidth with >50mW/nm spectral power density in this region. The measurement of the true extent of the supercontinuum source was limited by the wavelength range of the optical spectrum analyzer. Based on the short wavelength cutoff of the supercontinuum, we anticipate that the supercontinuum extends until a wavelength of ~2μm.

**Acknowledgement**. The authors would like to thank the Ministry of Electronics and Information Technology, Ministry of Science and Technology and Ministry of Human Resource development, Government of India.

**Funding** Science and Engineering Research Board (SB/S3/EECE/0149/2015) and the INSPIRE Program, Department of Science and Technology, India

**REFERENCES**

1. J. K. Ranka, R. S. Windeler, and A. J. Stentz, "Visible continuum generation in air–silica microstructure optical fibers with anomalous dispersion at 800 nm," Opt. Lett. **25**, 25-27 (2000).
2. H. Kawagoe, S. Ishida, M. Aramaki, Y. Sakakibara, E. Omoda, H. Kataura, and N. Nishizawa, "Development of a high power supercontinuum source in the 1.7 μm wavelength region for highly penetrative ultrahigh-resolution optical coherence tomography," Biomed. Opt. Express **5**, 932-943 (2014).
3. C. Amiot, A. Aalto, P. Ryczkowski, J. Toivonen, and G. Genty, "Cavity enhanced absorption spectroscopy in the mid-infrared using a supercontinuum source," Appl. Phys. Lett. **111**, 061103 (2017).
4. J. H. Lee, K. Lee, S. B. Lee, and C. H. Kim, "Extended-reach WDM-PON based on CW supercontinuum light source for colorless FP-LD based OLT and RSOA-based ONUs," Opt. Fiber Technol. **15**, 310-319 (2009).
5. G. P. Agrawal, *Nonlinear Fiber Optics, 4th ed.* (Academic, 2007).
6. J. C. Travers, A. B. Rulkov, B. A. Cumberland, S. V. Popov, and J. R. Taylor, "Visible supercontinuum generation in photonic crystal fibers with a 400W continuous wave fiber laser," Opt. Express **16**, 14435-14447 (2008).
7. J. M. Dudley, G. Genty, and S. Coen, "Supercontinuum generation in photonic crystal fiber," Rev. Mod. Phys. **78**, 1135-1184 (2006).
8. D. J. Richardson, J. Nilsson, and W. A. Clarkson, "High power fiber lasers: current status and future perspectives [Invited]," J. Opt. Soc. Am. B **27**, B63-B92 (2010).
9. B. A. Cumberland, J. C. Travers, S. V. Popov, and J. R. Taylor, "29 W High power CW supercontinuum source," Opt. Express **16**, 5954-5962 (2008).
10. B. H. Chapman, S. V. Popov, and R. Taylor, "Continuous Wave Supercontinuum Generation through Pumping in the Normal Dispersion Region for Spectral Flatness," IEEE Photonics Technol. Lett. 24(15), 1325 (2012).
11. A. K. Abeeluck and C. Headley, "Continuous-wave pumping in the anomalous- and normal-dispersion regimes of nonlinear fibers for supercontinuum generation," Opt. Lett. **30**, 61-63 (2005).
12. A. K. Abeeluck, C. Headley, and C. G. Jørgensen, "High-power supercontinuum generation in highly nonlinear, dispersion-shifted fibers by use of a continuous-wave Raman fiber laser," Opt. Lett. **29**, 2163-2165 (2004).
13. D. V. Skryabin, A. V. Yulin, "Theory of generation of new frequencies by mixing of solitons and dispersive waves in optical fibers," Phys. Rev. E **72**, 016619 (2005).
14. V. R. Supradeepa, Y. Feng, J. W. Nicholson, "Raman fiber lasers," J. Opt. **19**, 023001 (2017).
15. V. Balaswamy, S. Arun, G. Chayran, and V. R. Supradeepa, "All passive architecture for high efficiency cascaded Raman conversion," Opt. Express **26**, 3046-3053 (2018)
16. S. Arun, V. Choudhury, R. Prakash, and V. R. Supradeepa, "High Power, Tunable, L-Band (1.6micron Wavelength Region) Fiber Lasers," in *13th International Conference on Fiber Optics and Photonics*, OSA Technical Digest (online) (Optical Society of America, 2016), paper Tu3E.5.
17. S. Arun, V. Choudhury, V. Balaswamy, R. Prakash, and V. R. Supradeepa, "High power, high efficiency, continuous-wave supercontinuum generation using standard telecom fibers," Opt. Express 26, 7979-7984 (2018)
18. L. V. Kotov et al., "Submicrojoule femtosecond erbium-doped fibre laser for the generation of dispersive waves at submicron wavelengths, " Quantum Electron., vol. 44, no. 5, pp. 458–464, 2014.
19. R. Prakash, V. Choudhury, and V. R. Supradeepa, "High power, Continuous Wave, Supercontinuum Generation using Erbium-Ytterbium Co-doped Fiber Lasers," in *13th International Conference on Fiber Optics and Photonics*, OSA Technical Digest (online) (Optical Society of America, 2016), paper W3A.78.